\documentclass[prd,aps,eqsecnum,amsmath,floatfix,nofootinbib,preprint,tightenlines]{revtex4}

\usepackage{latexsym}
\usepackage{graphicx}
\usepackage{multirow}
\usepackage[dvipsnames]{xcolor}

\def\bibi{\bibitem}

\def\a{\alpha}
\def\b{\beta}
\def\c{\chi}
\def\d{\delta}
\def\e{\epsilon}                
\def\g{\gamma}

\def\j{\psi}

\def\l{\lambda}
\def\m{\mu}
\def\n{\nu}

\def\p{\pi}                     
\def\s{\sigma}                  
\def\t{\tau}

\def\z{\zeta}

\def\L{\Lambda}

\def\S{\Sigma}

\def\cl{{\cal L}}

\def\cbo{{\,\raise-.15ex\Sc [\,}}                       

\def\Sl#1{\rlap{\hbox{$\mskip 3 mu /$}}#1}      

\def\ddt#1{{\buildrel {\hbox{\LARGE .\kern-2pt.}} \over {#1}}}

\def\tr{{\rm tr}\,}

\def\half{{1\over 2}}

\def\ttl#1{{\it #1}}

\def\hd{\hat{\delta}}
\def\hg{\hat{g}}
\def\bj{\overline{\j}}
\def\ta{\tilde\a}

\begin{document}

\begin{boldmath}
\begin{center}
{\large\bf
Effective pion mass term and the trace anomaly
}\\[8mm]
Maarten Golterman$^a$ and Yigal Shamir$^b$\\[8 mm]
{\small
$^a$Department of Physics and Astronomy, San Francisco State University,\\
San Francisco, CA 94132, USA\\
$^b$Raymond and Beverly Sackler School of Physics and Astronomy,\\
Tel~Aviv University, 69978, Tel~Aviv, Israel}\\[10mm]
\end{center}
\end{boldmath}

\begin{quotation}
Recently, we developed an effective theory of pions and a light
dilatonic meson for gauge theories with spontaneously broken chiral symmetry
that are close to the conformal window.
The pion mass term in this effective theory
depends on an exponent $y$.
We derive the transformation properties under dilatations
of the renormalized fermion mass, and use this to rederive $y=3-\g_m^*$,
where $\g_m^*$ is fixed-point value of the mass anomalous dimension
at the sill of the conformal window.   This value for $y$
is consistent with the trace anomaly of the underlying near-conformal
gauge theory.
\end{quotation}

\newpage
\section{\label{intro} Introduction}
In Ref.~\cite{PP} we developed a low-energy effective action of pions
and a dilatonic meson, which are the pseudo Nambu-Goldstone bosons
for approximate chiral and scale symmetries, respectively, in
near-conformal gauge theories that still undergo dynamical chiral
symmetry breaking, and in which the scale symmetry is broken by the
trace anomaly. The effective theory is organized in terms of a systematic power
counting in
\begin{equation}
  p^2/\L^2 \ \sim \ m/\L
  \ \sim \ 1/N \ \sim \ |n_f-n_f^*| \ \sim \ \d \ ,
\label{pc}
\end{equation}
where $\d$ stands for the small expansion parameter.
As in the usual chiral lagrangian, $m$ is the fermion mass (we assume
a common mass for all flavors for simplicity), and $p^2$ is a shorthand
for the product of two external momenta, while $\L$ is the scale
associated with chiral symmetry breaking.
We invoke the Veneziano limit $N\to\infty$,
where the number of colors $N_c=N$ tends to infinity in proportion to
the number of fundamental representation flavors $N_f$ \cite{VZlimit}.
Here $n_f=N_f/N_c$, and $n_f^*$ is the critical value
where the conformal window is entered in the Veneziano limit,
\begin{equation}
  n_f^* = \lim_{N_c\to\infty} \frac{N_f^*(N_c)}{N_c} \ .
\label{nfstar}
\end{equation}
It is defined in terms of $N_f^*(N_c)$, which, in turn, is the smallest number
of flavors for which the $SU(N_c)$ theory is infrared conformal.

The effective action is constructed in terms of the usual chiral source field
$\c_{ij}(x)$, where $i,j=1,2,\ldots,N_f$ are flavor indices,
and the ``dilaton'' source field $\s(x)$, as well as effective fields
for the pions and the dilatonic meson.
After setting these sources equal
to their expectation values, $\c_{ij}(x)=m\d_{ij}$
and $\s(x)=0$, the leading-order lagrangian of the effective theory is
\cite{PP}
\begin{equation}
  \cl = \cl_\p + \cl_\t + \cl_m + \cl_d \ ,
\label{Leff}
\end{equation}
where
\begin{eqnarray}
  \cl_\p &=& \frac{f_\p^2}{4}\, e^{2\t}
            \tr(\partial_\m \S^\dagger \partial_\m \S) \ ,
\label{Lp}\\
  \cl_\t &=& \frac{f_\t^2}{2}\, e^{2\t} (\partial_\m \t)^2  \ ,
\label{Lt}\\
  \cl_m &=& -\frac{f_\p^2 B_\p m}{2} \, e^{y\t}
  \tr\Big( \S + \S^\dagger \Big) \ ,
\label{Lm}\\
  \cl_d &=& f_\t^2 B_\t \, e^{4\t} (c_0 + c_1\t) \ .
\label{Ld}
\end{eqnarray}
Here $\S(x)\in SU(N_f)$ is the usual non-linear field describing the pions,
while $\t(x)$ is the effective field of the dilatonic meson;
$f_\p$, $f_\t$, $B_\p$, $B_\t$ and $c_{0,1}$ are low-energy constants.

In Ref.~\cite{PP} we argued that the exponent $y$ in Eq.~(\ref{Lm}) is given by
\begin{equation}
  y = 3 - \g_m^* \ ,
\label{gammay}
\end{equation}
where $\g_m^*$ is the mass anomalous dimension at the sill of the
conformal window in the Veneziano limit.\footnote{%
More precisely, $\g_m^*=\g_m(\ta_*)$, where $\ta_*$ is infrared fixed-point
't~Hooft coupling at the sill of the conformal window.
}
Here we give a more complete
derivation of this result, and explore its relation to the trace anomaly.
This new derivation was motivated by the observation, made in Ref.~\cite{y3},
that $y$ may be determined by matching the divergences of the
dilatation current in the underlying and effective theories.\footnote{%
However, we disagree with the actual result for $y$ claimed in Ref.~\cite{y3}.}

The power counting~(\ref{pc}) hinges on the {\em assumption}
that, near the chiral symmetry breaking scale $\L$, the beta function
of the renormalized 't~Hooft coupling $\ta_r\equiv g^2N/(16\p^2)$ behaves like
\cite{PP,latt16}
\begin{equation}
  \b(\ta_r(\L)) = O(n_f-n_f^*) + O(1/N) \ .
\label{tbgN}
\end{equation}
This relation expresses the fact that the theory is
on the verge of developing an infrared attractive fixed point
(a situation that is sometimes referred to as ``emergent'' scale invariance).
The systematic expansion in $n_f-n_f^*$ derives from this central
dynamical assumption.

In order that the effective action will manifestly exhibit the expansion
in $n_f-n_f^*$ we have to choose the renormalization scale $\m$ such that
Eq.~(\ref{tbgN}) is applicable.  In other words, we need $\m\sim\L$;
we must renormalize the microscopic theory near
the scale where chiral symmetry breaking takes place.
Since the microscopic and the effective theories depend on the same set
of external sources, the mass parameter occurring in Eq.~(\ref{Lm}) is therefore
the renormalized mass, $m=m_r(\m)$, with the renormalization scale $\m$
chosen as above.  As we will show in Sec.~\ref{dilmass},
Eq.~(\ref{gammay}) is then a direct consequence
of the transformation properties of $m_r(\m)$ under dilatations.\footnote{%
We refer to Ref.~\cite{PP} for a general discussion of the dilatation transformation
properties of the effective theory,
both with and without the $\s(x)$ source field.
}
In Sec.~\ref{tracean} we explore the matching procedure proposed
in Ref.~\cite{y3}, finding that this procedure reproduces Eq.~(\ref{gammay}) as well.
Sec.~\ref{conc} contains our conclusions.

\section{\label{dilmass} Dilatation transformation of the renormalized mass}
We start from the bare lagrangian of the microscopic theory,
using dimensional regularization.
After a rescaling of the bare gauge and fermion fields by the
bare coupling $g_0$, the $d$-dimensional action is
\begin{equation}
  S = \int d^dx\, \frac{\m_0^{(d-4)}}{\hg_0^2} \, ( \cl_k + \cl_m ) \ ,
\label{bareSmu0glbl}
\end{equation}
where
\begin{eqnarray}
  \cl_k &=& \frac{1}{4} F_{\m\n}^a F_{\m\n}^a
  + \bj_i \Sl{D} \j_i \ ,
\label{bareLk}\\
  \cl_m &=& m_0 \bj_i\j_i \ ,
\label{Lm0}
\end{eqnarray}
and $m_0$ is the bare mass.
To expose the fact that $g_0$ is dimensionful for $d\ne4$ we substituted
\begin{equation}
  g_0 = \hg_0\, \m_0^{2-d/2} \ ,
\label{tg0}
\end{equation}
so that $\hg_0$ is dimensionless for any $d$.  As the only dimensionful
parameter in the massless bare action, $\m_0$ may be interpreted
as an ultraviolet cutoff scale.

A dilatation transformation acts on the fields and parameters occurring
in the bare lagrangian according to their canonical dimension,
\begin{subequations}
\label{Smu0glbl}
\begin{eqnarray}
  A_\m(x) &\to& \l A_\m(\l x) \ ,
\label{Smu0glbla}\\
  \j(x) &\to& \l^{3/2} \j(\l x) \ , \qquad
  \bj(x) \ \to \ \l^{3/2} \bj(\l x) \ ,
\label{Smu0glblb}\\
  m_0 &\to& \l m_0 \ ,
\label{Smu0glblc}\\
  \m_0 &\to& \l \m_0  \ ,
\label{Smu0glbld}\\
  \hg_0 &\to& \hg_0 \ .
\label{Smu0glble}
\end{eqnarray}
\end{subequations}
It is easy to check that the $d$-dimensional action is invariant
under this transformation. Here we take $m_0$ and $\m_0$
to be global spurions, with transformation rules that make the bare
action~(\ref{bareSmu0glbl}) invariant.\footnote{
The formulation in terms of spurion {\em fields},
and its relation with Eqs.~(\ref{bareSmu0glbl}) and~(\ref{Smu0glbl}),
will be discussed in Sec.~\ref{tracean} below.
For our purposes here the global spurions $m_0$ and $\m_0$ are sufficient.}

We next proceed to the renormalized parameters.
The renormalized coupling $g_r$ is defined as usual via
\begin{equation}
  Z_g(\e;g_r) g_r = \m^{-\e} g_0
  = (\m_0/\m)^{\e} \, \hg_0 \ ,
\label{mu0Zg}
\end{equation}
and the renormalized mass $m_r$ by
\begin{equation}
  m_0 = m_r \, Z_m(\e;g_r) \ .
\label{m0r}
\end{equation}
Here $\e=2-d/2$, and in the second equality of Eq.~(\ref{mu0Zg})
we have used Eq.~(\ref{tg0}).  We use a mass-independent renormalization scheme,
which implies that all $Z$ factors have a series expansion in $1/\e$
and in $g_r^2$.  It follows that the renormalized coupling itself,
as well as all the $Z$ factors, depend on $\m$ and $\m_0$ only through
their ratio $\m/\m_0$.  In particular, the renormalized mass satisfies
the renormalization-group equation
\begin{equation}
  \frac{\partial \log m_r}{\partial\log\m} = -\g_m
  = -\frac{\partial \log Z_m}{\partial \log\mu}
  = \frac{\partial \log Z_m}{\partial \log\mu_0} \ ,
\label{gamma}
\end{equation}
where in the last equality we have used that $Z_m=Z_m(\m/\m_0)$.

The dilatation transformation rule of the renormalized mass
is obtained as follows.  What needs to be calculated is the response
of the renormalized mass to the transformation~(\ref{Smu0glbl}),
which is applied to the bare theory while holding fixed
the physical scale represented by the renormalization scale $\m$.
Letting $m_0(\l)=\l m_0$ and $\m_0(\l)=\l \m_0$,
the transformation rule of $m_r=m_r(\l)$ under an infinitesimal dilatation
is obtained by differentiating Eq.~(\ref{m0r}) with respect to $\log\l$,
\begin{equation}
  \frac{1}{Z_m} \frac{\partial m_0}{\partial \log\l}
  = \frac{\partial m_r}{\partial \log\l}
    + \frac{m_r}{Z_m} \frac{\partial \m_0}{\partial \log\l}
          \frac{\partial Z_m}{\partial \m_0} \ .
\label{dmra}
\end{equation}
The derivative with respect to $\m_0$ can be traded with a derivative
with respect to $\m$ with the help of Eq.~(\ref{gamma}).  It follows that
\begin{equation}
  m_r = \frac{m_0}{Z_m}
  =  \frac{1}{Z_m} \frac{\partial m_0}{\partial \log\l}
  = \frac{\partial m_r}{\partial \log\l} - m_r \g_m \ ,
\label{dmrb}
\end{equation}
or
\begin{equation}
  \frac{\partial m_r}{\partial \log\l}
  = (1+\g_m) m_r \ .
\label{dmrc}
\end{equation}

As explained in the introduction,
in the effective theory we are expanding in $n_f-n_f^*$.
In the limit $n_f\nearrow n_f^*$ (and $N\to\infty$)
the 't~Hooft coupling behaves like  $\ta_r(\L)\to \ta_*$,
where $\ta_*$ is the location of the infrared fixed point
at the sill of the conformal window, in the Veneziano limit.
For small $|n_f-n_f^*|$, $\ta_r(\L)$ is close to $\ta_*$, and
$\g_m(\ta_r(\L))=\g_m^*$ up to corrections of order $n_f-n_f^*$.
In the leading-order effective action we thus have $\g_m=\g_m^*$.
Since this is a constant, we can integrate Eq.~(\ref{dmrc}) in closed form,
obtaining the transformation rule under a finite dilatation
\begin{equation}
  m_r \to \l^{1+\g_m^*} m_r \ .
\label{dmrg}
\end{equation}
The dilatation transformation rules of the effective fields are \cite{PP}
\begin{subequations}
\label{scleff}
\begin{eqnarray}
  \S(x) &\to& \S(\l x) \ ,
\label{scleffa}\\
  \t(x) &\to& \t(\l x) + \log \l \ .
\label{scleffb}
\end{eqnarray}
\end{subequations}
Using Eqs.~(\ref{dmrg}) and~(\ref{scleff}) it is easy to check that $\cl_m$
is invariant under dilatations if and only if $y$ is given by Eq.~(\ref{gammay}).
This is our main result.

\section{\label{tracean} Relation to the trace anomaly}
It was recently observed in Ref.~\cite{y3} that the relation of $y$
with the mass anomalous dimension may be inferred
via the following alternative procedure.  One first obtains
the divergence of the dilatation current $\partial_\m S_\m$ by
applying a suitable differential operator to the action.  This is done
separately in the (bare) microscopic theory and in the effective theory.
One then requires that {\em the same} differential operator will yield
$\partial_\m S_\m$ in both cases, following the general requirement
that correlation functions obtained by differentiating the partition functions
of the microscopic and of the effective theories will match.\footnote{
Note that $S_\m$ itself does not renormalize, because it is equal
to $x_\n T_{\m\n}$ with $T_{\m\n}$ the conserved energy-momentum tensor.}
In particular,
the same differential operator that yields $\partial_\m S_\m$ in the
microscopic theory should therefore also reproduce $\partial_\m S_\m$
at the level of the effective theory.
In Ref.~\cite{y3} it was claimed that the outcome of this consistency
requirement is that
$y=3$, which is in conflict with the value we derived in Sec.~\ref{dilmass}.
Here we will show that also this procedure leads to $y=3-\g_m^*$,
in disagreement with Ref.~\cite{y3}.   The key
point is that, before it can be applied to the effective theory,
the differential operator needs to be expressed in terms
of the renormalized mass.

We begin by coupling the bare action to local sources, which is done
by replacing Eq.~(\ref{bareSmu0glbl}) with
\begin{equation}
  S = \int d^dx\, \frac{(\m_0 e^{\s(x)})^{(d-4)}}{\hg_0^2} \,
  ( \cl_k + \cl_{src} ) \ ,
\label{bareSmu0}
\end{equation}
where now
\begin{equation}
  \cl_{src} = \half
  \Big( \c_{ij}\,\bj_i(1+\g_5)\j_j + \c^*_{ji}\,\bj_i(1-\g_5)\j_j \Big) \ .
\label{Lsrc}
\end{equation}
The dilatation transformation rules of the dynamical bare fields
remain the same as before, whereas the transformation rules
of the source fields and the parameters are given by
\begin{subequations}
\label{Smu0}
\begin{eqnarray}
  \c(x) &\to& \l \c(\l x) \ ,
\label{Smu0c}\\
  \s(x) &\to& \s(\l x) + \z \log \l \ ,
\label{Smu0d}\\
  \m_0 &\to& \l^{(1-\z)} \m_0  \ ,
\label{Smu0e}\\
  \hg_0 &\to& \hg_0 \ .
\label{Smu0f}
\end{eqnarray}
\end{subequations}
Notice the freedom to choose the parameter $\z$, which follows from
the redundancy between $\m_0$ and the constant mode of the $\s(x)$ field,
which we will denote as $\s_0$.

There are two variants of the matching procedure.
One can obtain $\partial_\m S_\m(x)$ via suitable differentiations
with respect to the local sources;
or one can obtain the integrated version $\int d^dx\, \partial_\m S_\m(x)$,
for which we may set the local sources to the constant values
$\s(x)=\s_0$, $\c_{ij}(x)=m_0\d_{ij}$.  Here we will choose
the second variant.

The integrated divergence $\int d^dx\, \partial_\m S_\m(x)$
is obtained by applying an infinitesimal dilatation
to the dynamical fields only.  Since $S$ is invariant when the
dilatation is applied to both fields and sources or parameters, it follows that
\begin{equation}
  \int d^dx\, \partial_\m S_\m = -\hd S \ ,
\label{dSspur}
\end{equation}
where the differential operator on the right-hand side is
\begin{equation}
  \hd = \z \frac{\partial}{\partial\s_0}\Bigg|_{m_0,\m_0}
  + (1-\z)\m_0 \frac{\partial}{\partial\m_0}\Bigg|_{m_0,\s_0}
  + m_0 \frac{\partial}{\partial m_0}\Bigg|_{\m_0,\s_0} \ ,
\label{dparz}
\end{equation}
and we have indicated explicitly which parameters are held fixed during
each differentiation.
Because the bare action depends only on $e^{\s_0}\m_0$,
the only change in Eqs.~(\ref{mu0Zg}) and~(\ref{m0r}) is that now $g_r$
and all the $Z$ factors are functions of the combination
\begin{equation}
  \frac{e^{\s_0}\m_0}{\m} \ .
\label{sm0m}
\end{equation}
It follows that the derivative $\partial/\partial\s_0$ is interchangeable
with $\partial/\partial\log\m_0$.
We thus have the alternative forms,
\begin{equation}
  \hd \ = \ \m_0 \frac{\partial}{\partial\m_0}\Bigg|_{m_0,\s_0}
  + m_0 \frac{\partial}{\partial m_0}\Bigg|_{\m_0,\s_0}
  \ = \ \frac{\partial}{\partial\s_0}\Bigg|_{m_0,\m_0}
  + m_0 \frac{\partial}{\partial m_0}\Bigg|_{\m_0,\s_0} \ ,
\label{dpar}
\end{equation}
which correspond to $\z=0$ and $\z=1$ in Eq.~(\ref{dparz}).

As explained in the introduction, the effective theory depends on
the renormalized mass $m_r$ (or, more generally, on the renormalized
chiral source $\c_r(x)$).  In order to be able to compare the action
of $\hd$ on the bare microscopic action and on the effective action,
we must first trade the bare mass for the renormalized mass.  One has
\begin{eqnarray}
  \m_0 \frac{\partial}{\partial\m_0}\Bigg|_{m_0,\s_0}
  &=& \m_0 \frac{\partial}{\partial\m_0}\Bigg|_{m_r,\s_0}
      + \m_0 \frac{\partial m_r}{\partial\m_0} \Bigg|_{m_0,\s_0}
        \frac{\partial}{\partial m_r}\Bigg|_{\m_0,\s_0}
\label{dmu0ren}\\
  &=& \rule{0ex}{4.5ex}
      \m_0 \frac{\partial}{\partial\m_0}\Bigg|_{m_r,\s_0}
      + \g_m m_r \frac{\partial}{\partial m_r}\Bigg|_{\m_0,\s_0} \ ,
\nonumber
\end{eqnarray}
where the last step follows from Eq.~(\ref{gamma}).\footnote{%
The term proportional to $\g_m$ was overlooked in Ref.~\cite{y3}.
}
In addition, Eq.~(\ref{m0r}) implies that
\begin{equation}
   m_0 \frac{\partial}{\partial m_0}\Bigg|_{\m_0}
   \ = \  m_r \frac{\partial}{\partial m_r}\Bigg|_{\m_0} \ ,
\label{dm0ren}
\end{equation}
when acting on the microscopic action.
Putting it together gives
\begin{subequations}
\label{dparren}
\begin{eqnarray}
  \hd &=& \m_0 \frac{\partial}{\partial\m_0}\Bigg|_{m_r,\s_0}
          + (1+\g_m) m_r \frac{\partial}{\partial m_r}\Bigg|_{\m_0,\s_0}
\label{dparrena}\\
  &=& \rule{0ex}{4.5ex}
      \frac{\partial}{\partial\s_0}\Bigg|_{m_r,\m_0}
      + (1+\g_m) m_r \frac{\partial}{\partial m_r}\Bigg|_{\s_0,\m_0} \ .
\label{dparrenb}
\end{eqnarray}
\end{subequations}

In the effective theory, the leading-order expressions for
the Noether current $S_\m$
and its divergence were calculated in Ref.~\cite{PP}.
As first observed in Ref.~\cite{y3}, the leading-order $\partial_\m S_\m$
may also be obtained as follows.  Starting from the leading-order
effective action $S_{\rm eff}$ which depends on the dilaton source $\s(x)$
and the renormalized chiral source $\c_{ij,r}(x)$
as detailed in Sec.~3.2 of Ref.~\cite{PP},
we let $\s(x)=\s_0$ and $\c_{ij,r}(x)=\d_{ij}m_r$,
and find that the (integrated) divergence of $S_\m$ is equal to
\begin{equation}
  \int d^4x\, \partial_\m S_\m = -\hd_{\rm eff} S_{\rm eff} \ ,
\label{dSeft}
\end{equation}
where
\begin{equation}
  \hd_{\rm eff} =
  \frac{\partial}{\partial\s_0}\Bigg|_{m_r} +
  (4-y) m_r \frac{\partial}{\partial m_r}\Bigg|_{\s_0} \ ,
\label{spureft}
\end{equation}
and $\s_0$ is set equal to zero in the end.
The requirement that the effective theory match the microscopic theory
thus implies that the differential operators $\hd$ and $\hd_{\rm eff}$
must be the same.  Comparing Eqs.~(\ref{dparrenb}) and~(\ref{spureft})
(and remembering that the effective theory does not depend explicitly
on $\m_0$) shows that this agreement will be reached provided that
\begin{equation}
  4-y=1+\g_m \ .
\label{4yg}
\end{equation}
Finally, taking the renormalization scale to be as described in the
introduction, we reproduce $\g_m=\g_m^*$ to leading order in the power counting,
and, with that, Eq.~(\ref{gammay}) as well.

\bigskip

The manipulations we have carried out in this section are closely related to the
original derivation of the trace anomaly in Ref.~\cite{CDJ},
which we will refer to as ``CDJ.''
At the starting point, CDJ introduces a parameter $a$
(called the loop expansion parameter), and multiplies the bare
lagrangian by $1/a$.  If we again rescale the bare fields as
we did in Sec.~\ref{dilmass}, so that the bare gauge coupling appears
as an overall factor $1/g_0^2$ in front of the lagrangian density,
we reach the equivalence
\begin{eqnarray}
  \frac{1}{ag_0^2}
  &\Leftrightarrow&
  \frac{e^{(d-4)\s_0}}{g_0^2}
  \ = \ \frac{(\m_0 e^{\s_0})^{d-4}}{\hg_0^2} \ ,
\label{equiv}
\end{eqnarray}
where the left-hand side refers to CDJ, and the right-hand side
to Eq.~(\ref{bareSmu0}) (with $\s(x)=\s_0$).  We already observed that
our $Z$ factors depend on $\m$, $\m_0$ and $\s_0$ only through
the variable~(\ref{sm0m}).  If we consider the dependence on $\hg_0$ as well,
our $Z$ factors depend only on the variable
\begin{equation}
   \frac{1}{\hg_0^2} \bigg(\frac{e^{\s_0}\m_0}{\m}\bigg)^{d-4}  \ .
\label{sm0mg}
\end{equation}
Correspondingly, CDJ observes that their $Z$ factors depend only on
\begin{equation}
  \frac{\m^{4-d}}{a g_0^2}  \ .
\label{cdj}
\end{equation}
At a key step in the argument,
CDJ then trades the derivatives with respect to $a$ (holding $\m$ fixed)
with derivatives with respect to the renormalization scale $\m$.
Evidently, what we have done is completely analogous, taking
derivatives with respect to $\m_0$ (or with respect to $\s_0$),
which are the variables that play the role of $a$ in our setting,
and trading them for derivatives with respect to $\m$.
This close correspondence has to exist, because,
as we have seen, acting on the bare action with the differential operator
$\hd$ generates the integral of $\partial_\m S_\m$.  In particular,
the term proportional to $\g_m$ in Eq.~(\ref{dparren}) corresponds directly
to the term proportional to $\g_m$ in the trace anomaly.

\section{\label{conc} Conclusion}
In this note we provided a more complete discussion of the
relation between the parameter $y$ in the pion-mass term of the
effective lagrangian~(\ref{Leff}) and the mass anomalous dimension $\g_m$,
thereby confirming the result already conjectured in Ref.~\cite{PP}.
We traced the incorrect result obtained in Ref.~\cite{y3} to the
difference between the behavior of
bare and renormalized sources under scale transformations.
We also pointed out that our result is required for consistency with
the expression for the trace anomaly in the underlying gauge theory
\cite{CDJ}.

In principle, the scale at which we renormalize the microscopic theory
can be chosen arbitrarily.
Imagine that we are very close to the gaussian fixed point at $g_0=0$,
by taking the renormalization scale $\m$ higher and higher.
We may then use the one-loop expression for the mass anomalous dimension,
so that $\g_m$ is linear in the renormalized 't~Hooft coupling $\ta_r(\m)$.
We see that $\g_m$ becomes
arbitrarily small if we take $\m$ arbitrarily large, and, eventually,
$y=3-\g_m(\ta_r(\m))$ will approach $y=3$.  This, of course, is merely
a reflection of the fact that the theory is asymptotically free.
However, if we want to define our effective theory~(\ref{Leff}) as the
leading term in a systematic expansion in $|n_f-n_f^*|\sim\d$,
we are forced to choose the renormalization scale near
the chiral symmetry breaking scale,
and, consequently, the difference $3-y=\g_m^*$ is of order one.

\vspace{2ex}
\noindent {\bf Acknowledgments}
\vspace{2ex}

We thank H.\ Suzuki for discussions.
This material is based upon work supported by the U.S. Department of
Energy, Office of Science, Office of High Energy Physics, under Award
Number DE-FG03-92ER40711.
YS is supported by the Israel Science Foundation
under grant no.~449/13.

\vspace{3ex}


\begin{thebibliography}{99}

\bibi{PP}
  M.~Golterman and Y.~Shamir,
  \ttl{Low-energy effective action for pions and a dilatonic meson,}
  Phys.\ Rev.\ D {\bf 94}, no. 5, 054502 (2016)
  [arXiv:1603.04575 [hep-ph]].

\bibi{VZlimit}
  G.~Veneziano,
  \ttl{Some Aspects of a Unified Approach to Gauge, Dual and Gribov Theories,}
  Nucl.\ Phys.\ B {\bf 117}, 519 (1976);
  \ttl{U(1) Without Instantons,}
  Nucl.\ Phys.\ B {\bf 159}, 213 (1979).

\bibi{y3}
  A.~Kasai, K.~i.~Okumura and H.~Suzuki,
  \ttl{Remark on the dilaton mass relation,}
  arXiv:1609.02264 [hep-lat].

\bibi{latt16}
  M.~Golterman and Y.~Shamir,
  \ttl{Effective field theory for pions and a dilatonic meson,}
  PoS LATTICE {\bf 2016}, 205 (2016)
  [arXiv:1610.01752 [hep-ph]]

\bibi{CDJ}
  J.~C.~Collins, A.~Duncan and S.~D.~Joglekar,
  \ttl{Trace and Dilatation Anomalies in Gauge Theories,}
  Phys.\ Rev.\ D {\bf 16}, 438 (1977).

\end{thebibliography}
\end{document}